\documentclass[aps,prc,preprint,tightenlines,showpacs,endfloats,byrevtex,%
nofootinbib,superscriptaddress]{revtex4}

\usepackage{epsfig,bm}

\begin{document}

\preprint{nucl-th/0205065} 

%%%%%%%%%%%%%%%%%%%%% Title %%%%%%%%%%%%%%%%%%%%%%

\title{Form-factor dependence of the $\bm{J/\psi}$ dissociation cross
sections \\ in meson exchange model}

%%%%%%%%%%%%%%%%%%%% Authors %%%%%%%%%%%%%%%%%%%%%

\author{Yongseok Oh}%
\email{yoh@phya.yonsei.ac.kr}

\author{Taesoo Song}%
\email{tssong@phya.yonsei.ac.kr}

\author{Su Houng Lee}%
\email{suhoung@phya.yonsei.ac.kr}

\affiliation{Institute of Physics and Applied Physics, Department of
Physics, Yonsei University, Seoul 120-749, Korea}

\author{Cheuk-Yin Wong}%
\email{wongc@ornl.gov}
\affiliation{Physics Division, Oak Ridge National Laboratory, Oak Ridge,
Tennessee 37831}

%%%%%%%%%%%%%%%%%%%% Abstract %%%%%%%%%%%%%%%%%%%%%

\begin{abstract}
The $J/\psi$ dissociation by $\pi$ and $\rho$ mesons is examined in
meson exchange model and compared to the quark interchange model.
We found that the main difference between the predictions of the
two models could be resolved by the form factors and the contact
interactions in meson exchange model calculations.
By adopting covariant form factors and adjusting the four-point
couplings, we found that the meson exchange model could give similar
predictions with the quark interchange models not only for the
magnitudes but also for the energy dependence of the low-energy
dissociation cross sections of the $J/\psi$ by $\pi$ and $\rho$ mesons.
Our finding suggests a way to understand one of the discrepancies among
the existing model predictions on the $J/\psi$ dissociation cross
sections by light hadrons.
\end{abstract}


\maketitle

%%%%%%%%%%%%%%%%%%%% Text %%%%%%%%%%%%%%%%%%%%%

The $J/\psi$ suppression has attracted much attention as a possible
signal of quark-gluon plasma (QGP) formation \cite{MS86}.
The recently observed anomalous $J/\psi$ suppression in Pb-Pb
collisions by the NA50 Collaboration \cite{NA50} was interpreted as a
strong indication of the production of a deconfined quark-gluon phase
in the collisions.
However, $J/\psi$ suppression is already present in hadron-nucleus
collisions and it is very important to disentangle the absorption of the
$J/\psi$ by nucleons and comovers from the measured data before one
makes a definitive conclusion on the formation of QGP \cite{Vogt99}.
At present, the observed $J/\psi$ suppression can be explained either
by the suppression due to the formation of QGP
\cite{NS98,Wong96,BO96a-BDO00} or by the suppression by comovers
\cite{GV96,CK97,ACF99-CFK00,STST00} with a sufficiently large 
dissociation cross sections by hadrons.
Therefore, it is necessary to understand the $J/\psi$ dissociation
cross sections by hadrons, which are the basic inputs in the comovers
approaches.

The dissociation process of the $J/\psi$ by hadrons has been considered
in several approaches, but the predicted cross sections show very
different energy dependence and magnitudes near threshold.
Hence careful analyses of the existing model calculations are
indispensable to understand their limitations and expected corrections.

Using the method of perturbative QCD, Peskin and Bhanot \cite{Pesk79-BP79}
and later Kharzeev and Satz \cite{KSa94} estimated the scattering cross
sections of the $J/\psi$ with light hadrons, which was recently 
rederived using the QCD factorization combined with the Bethe-Salpeter
amplitude \cite{OKL02-LO02}.
The pQCD result was recently improved by including the finite target 
mass corrections \cite{KSSZ96,AGGA02} and the relativistic correction 
to the phase space of the reactions \cite{OKL02-LO02}.
The estimated dissociation cross sections of the $J/\psi$ by light
hadrons depend on the gluon distribution function of the hadron.
As for the dissociation cross section by the nucleon, it is in the order
of $\mu$b near threshold and saturates to a few mb at higher energy.
However, the higher-twist effects are expected to be non-trivial and
should be properly accounted for \cite{OKL02-LO02}.

Martins {\em et al.\/} \cite{MBQ95} estimated the $J/\psi$
dissociation cross sections using the quark interchange model of Barnes
and Swanson \cite{BS92}.  
The calculation has been improved by Wong and collaborators by
employing a more realistic confining potential \cite{WSB00b-02}.
The estimated dissociation cross section of $\pi + J/\psi$ has a peak
at around $1$ mb and that of $\rho + J/\psi$ is in the order of several
mb near threshold.
The recent calculation using QCD sum rules \cite{NNMK02} seems to favor
the quark interchange model but shows different energy dependence.

Another approach to $J/\psi$ scattering with light hadrons is to use
effective meson Lagrangian
\cite{MMu98,Hagl00,LK00,OSL01,HG01,STT01,NNR01,LKL02,IKBB01}.
In this approach, one assumes SU(4) flavor symmetry to classify the
possible interactions of the $J/\psi$ by light mesons and fixes the 
necessary couplings either by phenomenology or by model calculations.
Matinyan and M{\"u}ller considered $t$-channel $D$-meson exchanges
only and found small values for the cross sections of the $J/\psi$
dissociation by $\pi$ and $\rho$.
However the small mass difference between the $D$ and the $D^*$ implies
that the $D^*$-meson exchange should  also be important.
Indeed, it was found by Haglin \cite{Hagl00} and Lin and Ko \cite{LK00}
that such additions increase the cross sections to several mb.
In Ref. \cite{OSL01}, Oh {\em et al.\/} have improved this model by
introducing anomalous parity interactions such as $D^* D^* \pi$ whose
coupling is related to the $D^* D \pi$ interaction by heavy quark
symmetry.
This opens new channels and additional mechanisms for the $J/\psi$
dissociation processes and it was found to decrease the
cross sections by up to about 50\% near threshold.

However, the resulting cross sections are different from those of
perturbative QCD and quark interchange models not only for the
magnitudes but also for energy dependence.
It should be also mentioned that the cross sections in the meson
exchange model depend strongly on the employed form factors which
could not be justified {\em a priori\/}.
Therefore, proper choice for the form factors is very crucial to
obtain more reliable model predictions since the range of heavy 
meson exchange is much smaller than the sizes of the initial hadrons.
It is especially important in this case, because the kinematical point
at which the couplings are fixed can be very different from the point
at which the couplings are actually used in the calculation.  
One example is the $g_{D^*D\pi}^{}$ coupling and its contribution to
the $D^*$ exchange in the $\pi + J/\psi \rightarrow D^* + \bar{D}$
process at high energy.
The couplings are determined at the on-shell point of each particle,
but in the exchange diagram at high energy the exchanged $D^*$ becomes
highly spacelike.
This means that a proper covariant form factor has to be used to
properly take the off-shell-ness of the exchanged particles into
account. 
In this paper, we show that the main difference between the meson
exchange model and the quark interchange model stems from the choice
of the form factors for the vertices in meson exchange model calculations.
In principle, the form factors and their cutoff parameters should be
determined by fitting the experimental data.
But because of the lack of any experimental data on $J/\psi$
dissociation processes, we compare our results with those of the quark
interchange model of Ref. \cite{WSB00b-02}.
By employing covariant form factors, we will show that meson exchange model
can mostly reproduce the characteristic features of the quark interchange
model predictions for the cross sections of $\pi(\rho) + J/\psi$.

We use the effective Lagrangian and the conventions adopted by
Ref. \cite{OSL01}, which reads
\begin{eqnarray}
{\mathcal{L}}_{D^* D \pi} &=& i g_{D^*D\pi}^{} \left( D_\mu^* \partial^\mu
\pi \bar{D} - D \partial^\mu \pi \bar{D}^*_\mu \right),
\nonumber \\
\mathcal{L}_{\psi DD} &=&
i g_{\psi DD}^{} \psi_\mu \left( \partial^\mu D
\bar{D} - D \partial^\mu \bar{D} \right),
\nonumber \\
{\mathcal{L}}_{\psi D^* D^*} &=& -i g_{\psi D^* D^*}^{} \Bigl\{
\psi^\mu \left( \partial_\mu D^{*\nu} \bar{D}_\nu^* - D^{*\nu}
\partial_\mu \bar{D}_\nu^* \right) 
\nonumber \\ && \mbox{}
+ \left( \partial_\mu \psi_\nu D^{*\nu} - \psi_\nu \partial_\mu D^{*\nu}
\right) \bar{D}^{*\mu}
\nonumber \\ && \mbox{}
+ D^{*\mu} \left( \psi^\nu \partial_\mu \bar{D}^*_\nu - \partial_\mu
\psi_\nu \bar{D}^{*\nu} \right)
\Bigr\},
\nonumber \\
{\mathcal{L}}_{\psi D^* D\pi} &=& g_{\psi D^* D\pi}^{} \psi^\mu \left( D
\pi \bar{D}^*_\mu + D^*_\mu \pi \bar{D} \right),
\nonumber \\
{\mathcal{L}}_{DD\rho} &=& i g_{DD\rho}^{} \left( D \rho^\mu \partial_\mu
\bar{D} - \partial_\mu D \rho^\mu \bar{D} \right),
\nonumber \\
{\mathcal{L}}_{D^*D^*\rho} &=& i g_{D^*D^*\rho}^{}  \Bigl\{
\partial_\mu D^*_\nu \rho^\mu \bar{D}^{*\nu} - D_\nu^* \rho_\mu
\partial^\mu \bar{D}^{*\nu}
\nonumber \\ && \mbox{}
+ \left( D^{*\nu} \partial_\mu \rho_\nu - \partial_\mu D_\nu^* \rho^\nu
\right) \bar{D}^{*\mu}
\nonumber \\ && \mbox{}
+ D^{*\mu} \left( \rho^\nu \partial_\mu \bar{D}^*_\nu - \partial_\mu
\rho_\nu \bar{D}^{*\nu} \right) \Bigr\},
\nonumber \\
{\mathcal{L}}_{\psi DD\rho} &=& -g_{\psi DD\rho}^{} \psi^\mu D \rho_\mu
\bar{D},
\nonumber \\
{\mathcal{L}}_{\psi D^* D^* \rho} &=& g_{\psi D^* D^* \rho}^{} \psi_\mu
\left( 2 D^{*\nu} \rho^\mu \bar{D}^*_\nu - D^{*\nu} \rho_\nu
\bar{D}^{*\mu} \right.
\nonumber \\ && \left.\mbox{} - D^{*\mu} \rho^\nu \bar{D}^*_\nu \right),
\nonumber \\
{\mathcal{L}}_{D^* D^* \pi} &=& -g_{D^* D^* \pi}^{}
\varepsilon^{\mu\nu\alpha\beta} \partial_\mu D^*_\nu \pi \partial_\alpha
\bar{D}^*_\beta,
\nonumber \\
{\mathcal{L}}_{\psi D^* D} &=& -g_{\psi D^* D}^{}
\varepsilon^{\mu\nu\alpha\beta} \partial_\mu \psi_\nu \left(
\partial_\alpha D^*_\beta \bar{D} + D \partial_\alpha \bar{D}^*_\beta
\right),
\nonumber \\
{\mathcal{L}}_{\psi DD \pi} &=& -i g_{\psi DD \pi}^{}
\varepsilon^{\mu\nu\alpha\beta} \psi_\mu \partial_\nu D \partial_\alpha
\pi \partial_\beta \bar{D},
\nonumber \\
{\mathcal{L}}_{\psi D^* D^* \pi} &=& -i g_{\psi D^* D^* \pi}^{}
\varepsilon^{\mu\nu\alpha\beta} \psi_\mu D^*_\nu \partial_\alpha \pi
\bar{D}^*_\beta
\nonumber \\ && \mbox{}
- i h_{\psi D^* D^* \pi}^{} \varepsilon^{\mu\nu\alpha\beta} \partial_\mu
  \psi_\nu D_\alpha^* \pi \bar{D}_\beta^*,
\nonumber \\
{\mathcal{L}}_{D^* D \rho} &=& -g_{D^* D \rho}^{}
\varepsilon^{\mu\nu\alpha\beta} \left( D \partial_\mu \rho_\nu
\partial_\alpha \bar{D}^*_\beta + \partial_\mu D^*_\nu \partial_\alpha
\rho_\beta \bar{D} \right),
\nonumber \\
{\mathcal{L}}_{\psi D^* D \rho} &=& i g_{\psi D^* D \rho}^{}
\varepsilon^{\mu\nu\alpha\beta} \psi_\mu \left( \partial_\nu D
\rho_\alpha \bar{D}^*_\beta + D^*_\nu \rho_\alpha \partial_\beta \bar{D}
\right)
\nonumber \\ && \mbox{}
- i h_{\psi D^* D \rho}^{} \varepsilon^{\mu\nu\alpha\beta} \psi_\mu
  \left( D \rho_\nu \partial_\alpha \bar{D}_\beta^* - \partial_\nu
D^*_\alpha \rho_\beta \bar{D} \right),
\nonumber \\
\end{eqnarray}
where $\pi = \bm{\tau} \cdot \bm{\pi}$, $\rho = \bm{\tau} \cdot
\bm{\rho}$ with $\varepsilon_{0123} = +1$.
The charm meson iso-doublets are defined as
$\bar{D}^T = \left( \bar{D}^0 ,  D^- \right)$,
$D = \left( D^0 , D^+ \right)$, etc.
The details on the derivation of the effective Lagrangian and the
determination of the couplings are explained in Refs.
\cite{Hagl00,LK00,OSL01} and will not be repeated here.
The only difference with the couplings of Ref. \cite{OSL01} lies on
the $g^{}_{D^* D \pi}$.
In Ref. \cite{OSL01}, the QCD sum rule prediction \cite{BBKR95} for the
coupling, i.e., $g^{}_{D^* D \pi} = 8.8$, was used since only the upper
limit of the $D^* \to D \pi$ decay width was known \cite{PDG00}.
But the recent measurement of $\Gamma(D^*)$ by the CLEO Collaboration
\cite{CLEO02} allows one to determine the coupling constant rather
precisely,
\begin{equation}
g_{D^* D \pi}^{} = 12.6 \pm 1.4,
\label{cleo}
\end{equation}
which is then used to determine the values for $g_{D^*D^*\pi}^{}$,
$g_{\psi D^* D \pi}$, $g^{}_{\psi DD\pi}$, $g^{}_{\psi D^* D^* \pi}$,
and $h^{}_{\psi D^* D^* \pi}$ according to the rules explained in
Ref. \cite{OSL01}.
Note that our convention for $g^{}_{D^* D \pi}$ is different
from that of Refs. \cite{BBKR95,CLEO02} by $1/\sqrt{2}$.

\begin{figure}[t]
\begin{center}
\epsfig{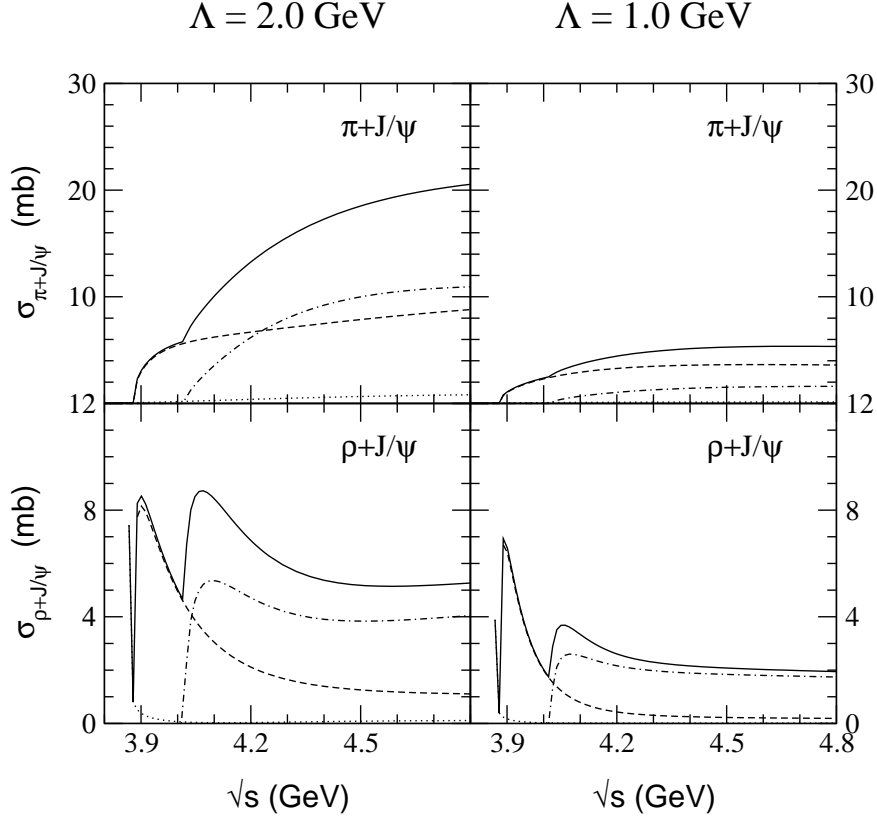}
\end{center}
\caption{Cross sections for $\pi + J/\psi$ and $\rho + J/\psi$ with
the form factors in Eq.~(\ref{FF-LK}) with $\Lambda = 2$ GeV (left panel)
and $1$ GeV (right panel).
The dotted lines are for $J/\psi + \pi(\rho) \to D+\bar{D}$, the dashed
lines for $J/\psi + \pi(\rho) \to D+\bar{D}^*$, $D^* + \bar{D}$, and the
dot-dashed lines for $J/\psi + \pi(\rho) \to D^* + \bar{D}^*$.
The solid lines are the sums.}
\label{fig:compall}
\end{figure}

\begin{figure}[t]
\centering
\epsfig{file=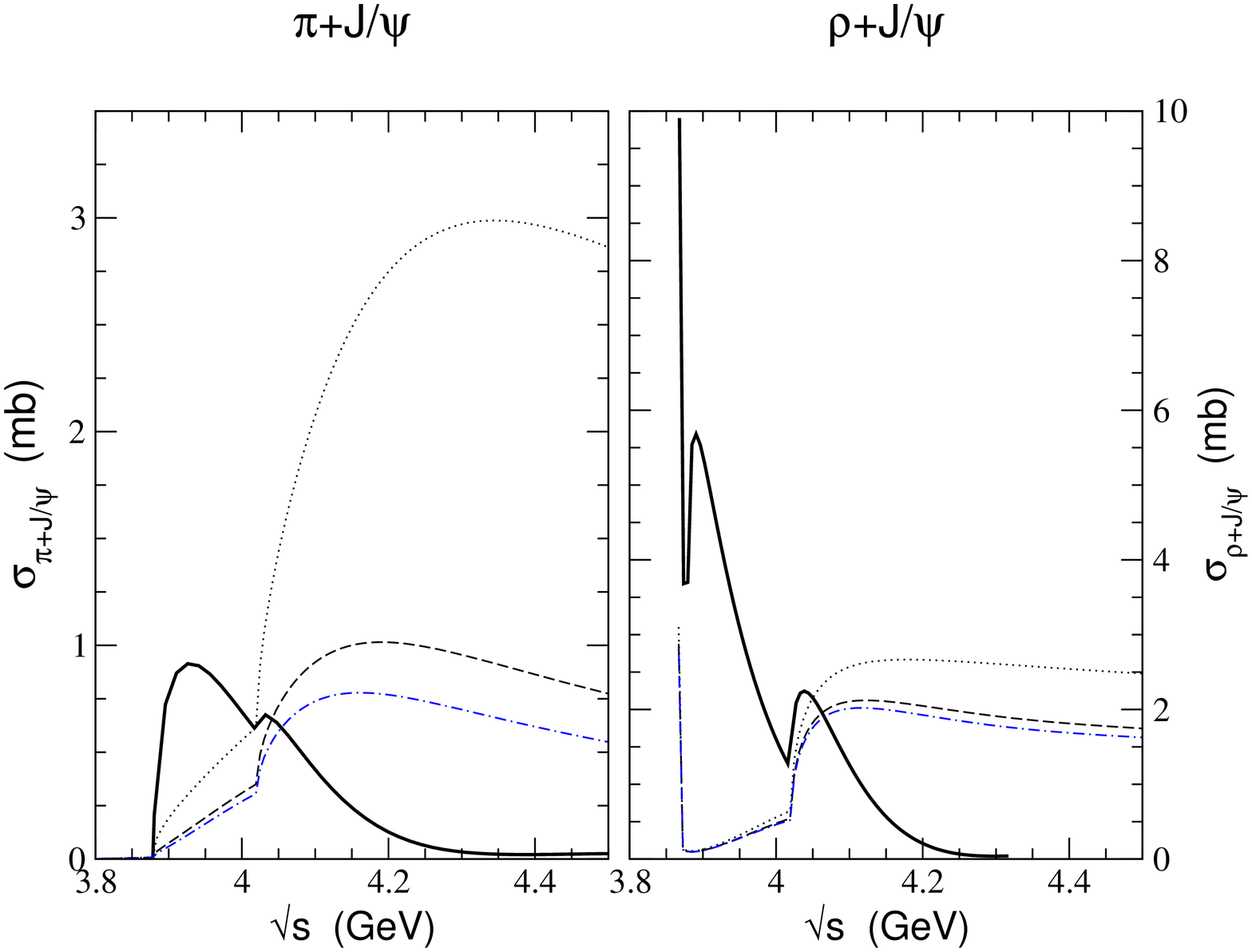, width=0.7\hsize,angle=-90}
\caption{Cross sections for $\pi + J/\psi$ and $\rho + J/\psi$ in model
(I).
The form factor (\ref{FF-PJ}) is used with $\Lambda = 1.8$ GeV for $\pi
+ J/\psi$ and $\Lambda = 2.3$ GeV for $\rho + J/\psi$.
The dotted, dashed, and dot-dashed lines are obtained with $n=1$, $5$,
and $20$, respectively.
The solid line is the quark interchange model result of Ref.
\cite{WSB00b-02}.}
\label{fig:model-1}
\end{figure}

\begin{figure}[t]
\centering
\epsfig{file=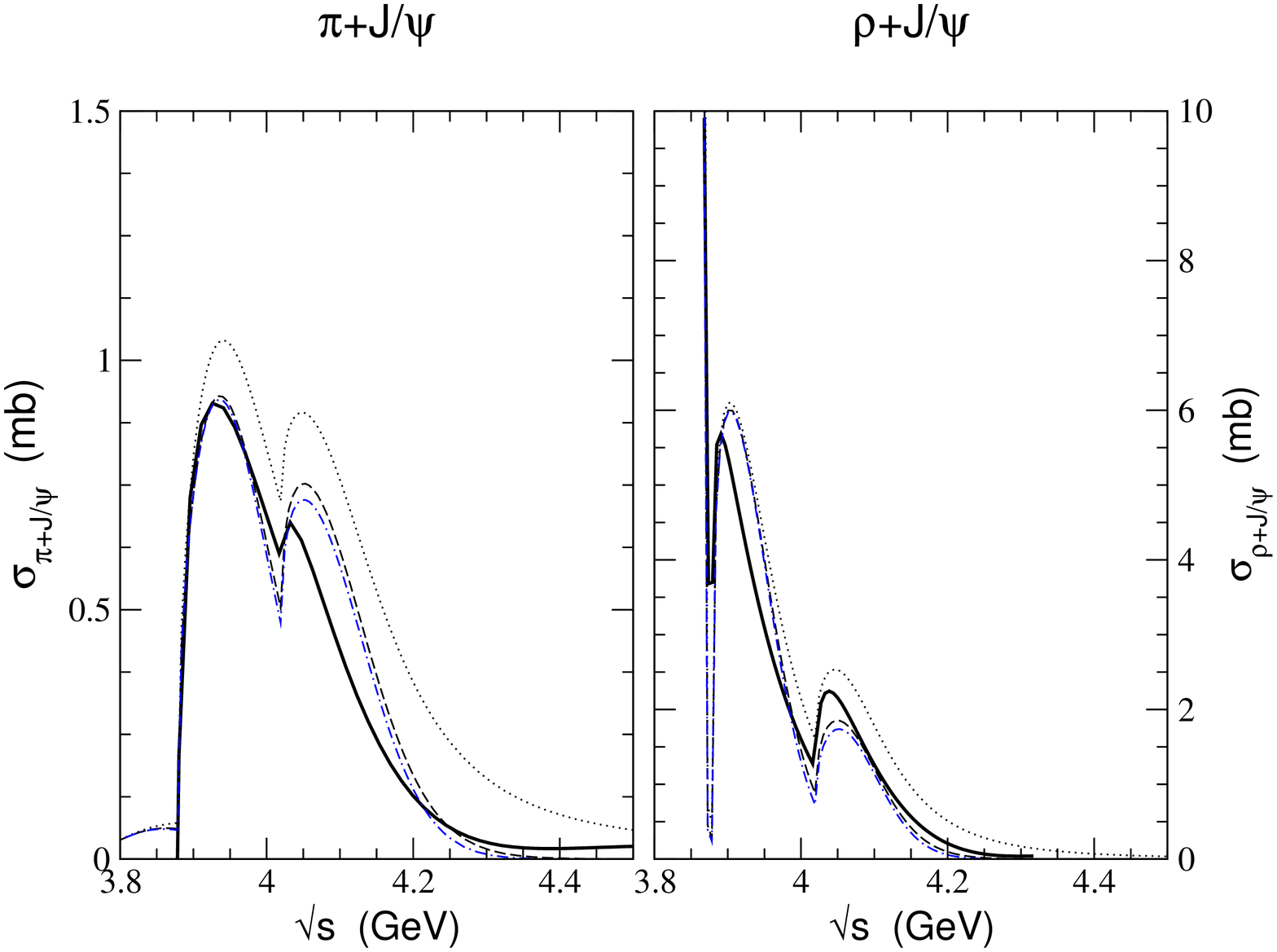, width=0.7\hsize,angle=-90}
\caption{Cross sections for $\pi + J/\psi$ and $\rho + J/\psi$ in model
(II).
The form factors (\ref{FF-PJ}) and (\ref{FF-PJ4}) are used 
with $\Lambda = 1.2$ GeV for $\pi
+ J/\psi$ and $\Lambda = 1.0$ GeV for $\rho + J/\psi$.
Notations are the same as in Fig.~\ref{fig:model-1}.}
\label{fig:model-2}
\end{figure}

Then the $J/\psi$ dissociation processes by $\pi$ and $\rho$ mesons
can be calculated from the $t$- and $u$-channel diagrams shown in
Figs.~1 and 2 of Ref. \cite{OSL01}, which are obtained by respecting the
OZI rule.%
\footnote{One may have $s$-channel diagrams by including the OZI-violating
interaction, $\mathcal{L}_{\psi\rho\pi} = g_{\psi\rho\pi}^{}
\varepsilon^{\mu\nu\alpha\beta} \partial_\mu \psi_\nu \mbox{Tr}\,
(\partial_\alpha \rho_\beta \pi )$, where $|g_{\psi\rho\pi}^{}| = 1.07
\times 10^{-3}$ GeV$^{-1}$ from the decay width of $J/\psi \to \rho\pi$
\cite{PDG00}. It allows $s$-channel diagrams ($\rho$ and $\pi$
exchanges) for $\pi+J/\psi \to D+\bar{D}, D^* + \bar{D}, D^* + \bar{D}^*$
and $\rho+J/\psi \to D^* + \bar{D}, D^*+\bar{D}^*$.
However we found that such diagrams give the cross sections at the order
of pb or fb and hence suppressed because of the small value of the
coupling constant $g_{\psi\rho\pi}^{}$.}
In Ref. \cite{OSL01}, the form factors were chosen to be
\begin{equation}
F_3(r) = \frac{\Lambda^2}{\Lambda^2 + r^2}, \qquad
F_4(r) = \frac{\Lambda^2}{\Lambda^2 + {\bar r}^2}
         \frac{\Lambda^2}{\Lambda^2 + {\bar r}^2}
\label{FF-LK}
\end{equation}
following Ref. \cite{LK00}, where $r^2 = ({\bf p}_1 - {\bf p}_3)^2$ or
$( {\bf p}_2 - {\bf p}_3)^2$ and $\bar{r}^2 = [ ({\bf p}_1 - {\bf
p}_3)^2 + ({\bf p}_2 - {\bf p}_3)^2]/2$.
Here $F_3(r)$ and $F_4(r)$ are the form factors for the three-point
and four-point vertices, respectively.
$p_1$ and $p_2$ are the $J/\psi$ and initial light meson ($\pi$ or
$\rho$) momentum, respectively, while $p_3$ and $p_4$ are the momenta
of the final state mesons.
The cutoff parameters may take different values for different vertices.
However, because of the paucity of experimental information, we use
the same cutoff parameters for all the vertices in this exploratory study
in order to investigate the dependence of the cross sections on
the cutoff parameters.
The cross sections thus obtained are shown in Fig.~\ref{fig:compall}, where
the numerical errors committed in Ref.~\cite{OSL01} are also corrected.
This evidently shows that the predictions for the magnitudes and the
energy dependence of the cross sections are quite different from those of
the quark interchange model \cite{WSB00b-02}, which are shown as solid
lines in Figs.~\ref{fig:model-1} and~\ref{fig:model-2}.

However, the form factors in Eq. (\ref{FF-LK}) are not Lorentz-invariant
and does not fully take into account the off-shell-ness of the exchanged 
particle.  
For example, because it depends on the three-momentum squared, they are
not normalized to be $1$ even when the exchanged particles are on-mass
shell.  
In the other extreme, it can be $1$ even for  particles far off-shell as 
long as they are at rest.
This may be a good starting approximation for very low energy reactions,
but we expect much corrections in $J/\psi$ dissociation processes
especially for large energies since the exchanged $D$ and $D^*$ mesons
will be {\em highly off-shell\/}.
A similar conclusion was drawn by Friman {\em et al.\/} \cite{FLK99}
by showing that the form factor (\ref{FF-LK}) would overestimate the
contribution of the $\Delta(1232)$ in the calculation for the vector
meson spectral densities in nuclear matter.
(See also Ref. \cite{PLM01}.)
Therefore, instead of the form factors in (\ref{FF-LK}), we will employ
the covariant form factor in (\ref{FF-PJ}) suggested by Pearce and
Jennings \cite{PJ91}, to study the dependence of the $J/\psi$
dissociation cross section on the form factor:
\begin{equation}
F(p^2) = \left( \frac{n \Lambda^4}{n \Lambda^4 + (p^2 - M_{\rm ex}^2)^2}
\right)^n. 
\label{FF-PJ}
\end{equation}
Here, $p$ is the four-momentum of the exchanged particle of mass
$M_{\rm ex}$.
(Other forms for the form factors are suggested in Refs.
\cite{HG01,IKBB01}.)
The function $F(p^2)$ is normalized to $1$ when the particles are 
on-shell ($p^2 = M_{\rm ex}^2$) and becomes smaller when they are
off-shell.
In the extreme limit $n \to \infty$, $F(p^2)$ approaches a Gaussian
in $(p^2 - M_{\rm ex}^2)$ with width $\Lambda^2$ \cite{PJ91}.
Here, we vary the cutoff $\Lambda$ and $n$ and compare the resulting
cross sections to the quark interchange model.

As for the models for $J/\psi$ dissociation processes, we consider two
cases.
In model (I), we include the meson-exchange diagrams with three-point
vertices only by discarding the four-point contact terms,
since their couplings are not well-determined.
The relevant diagrams can be identified from Figs.~1 and 2 of
Ref. \cite{OSL01}.
Given in Fig.~\ref{fig:model-1} are the results for $\pi + J/\psi$ (left
panel) and $\rho + J/\psi$ (right panel).
The form factor (\ref{FF-PJ}) is multiplied to each vertex with $\Lambda
= 1.8$ GeV for $\pi + J/\psi$ and $\Lambda = 2.3$ GeV for $\rho +
J/\psi$.
The dotted, dashed, and dot-dashed lines are obtained with $n=1$, $5$,
and $20$, respectively.
Our results show that the cross sections converge with increasing $n$.
As can be seen in Fig.~\ref{fig:model-1}, we find that the meson exchange
model with three-point vertices does not agree well with the quark
interchange model, especially for the energy dependence of the cross
sections.
Although our form factor at large $n$ should be related to the Gaussian
nature of the hadron wavefunctions in the quark model, our cross section 
for the process $\pi + J/\psi \to D^* + \bar{D}$ is different from that 
obtained in Ref. \cite{IKBB01}, which also employs exponential form
factors, motivated from the hadron wavefunctions, but which are
complicated functions of both $s$ and $t$ (or $u$).
Also, the form factors used in Ref. \cite{IKBB01} still do not satisfy the
on-shell condition.

In Fig. \ref{fig:model-2}, we consider model (II), which includes the
four-point contact diagrams in addition to the meson exchange diagrams.
Since the four-point couplings are only poorly known, we treat the
four-point couplings as free parameters and try to find a best fit to
the quark interchange model results.
For the form factors, we use
\begin{equation}
F(p^2) = \left( \frac{n \Lambda^4}{n \Lambda^4 + [(p_1+p_2)^2 -
(M_3 + M_4)^2]^2}
\right)^n
\label{FF-PJ4}
\end{equation}
for the contact terms assuming an internal structure of the couplings,
where $M_3$ and $M_4$ are the final state meson masses.
For the three-point vertices, the form factor (\ref{FF-PJ}) is used as in
model (I).
The results shown in Fig. \ref{fig:model-2} are obtained with $\Lambda =
1.2$ GeV for $\pi + J/\psi$ and $\Lambda = 1.0$ GeV for $\rho + J/\psi$.
As in Fig.~2, the  dotted, dashed, and dot-dashed lines are obtained 
with $n=1$, $5$, and $20$, respectively and the solid lines are from the
quark interchange model.
In this model, we found that the meson exchange terms (three-point vertices)
are overwhelmed by the contact terms (four-point vertices) such that their
contributions are found to be suppressed.
The coupling constants used in Fig. \ref{fig:model-2} are
$g_{\psi D D \pi}^{} = 23$ $(23)$,
$g_{\psi D^* D \pi}^{} = 34$ $(49)$,
$g_{\psi D^* D^* \pi}^{} = h_{\psi D^* D^* \pi}^{} = 7$ $(55)$,
$g_{\psi D D \rho}^{} = 39$ $(39)$,
$g_{\psi D^* D \rho}^{} = h_{\psi D^* D \rho}^{} = 17$ $(22)$, and
$g_{\psi D^* D^* \rho}^{} = 27$ $(19)$,
where the values in the parentheses are determined as in Ref. \cite{OSL01}
with $g_{D^* D\pi}^{}$ of Eq. (\ref{cleo}).
The $\pi + J/\psi \to D + \bar{D}$ and $\rho + J/\psi \to D + \bar{D}$
processes are suppressed compared to the other processes and we do not
vary the corresponding couplings.
The resulting magnitudes of the cross sections are strongly
dependent on the couplings and Fig.~\ref{fig:model-2} shows that the
quark interchange model would prefer model (II).
The big difference of the couplings from those in the parentheses would
imply the strong violation of the SU(4) symmetry relations used in
previous meson exchange model calculations \cite{OSL01}.
Therefore it is very important to estimate the four-point couplings and the
corresponding form factors within the quark models.

In summary, we have reinvestigated the $J/\psi$ dissociation cross sections
by $\pi$ and $\rho$ in effective Lagrangian approach.
We found that the quark interchange model predictions can be explained
by model (II) of meson exchange model, where the contact terms are
dominant.
Although this does not prove the predictions of the quark interchange
models on the $J/\psi$ dissociation processes, our finding suggests that
most of the discrepancies between the two models can be resolved by
choosing appropriate form factors and  strength for the contact terms,
which should be closely related to the hadron wavefunctions and interquark
forces in the quark model.
In addition, for a more detailed comparison between the two models, 
it is necessary to include the contributions from the exchange of 
higher resonances, such as the axial vector $D_1(2420)$, in the 
meson exchange model.
This is so because the exchanged quark-antiquark pairs in quark
interchange model implicitly contain all possible meson states.  

%\acknowledgments

This work was supported in part by the Brain Korea 21 project of Korean
Ministry of Education, KOSEF under Grant No. 1999-2-111-005-5, and
the Korean Ministry of Education under Grant No. 2000-2-0689.
This work was also supported in part by the Division of Nuclear Physics,
U. S. Department of Energy and the Laboratory Directed Research and
Development Program of Oak Ridge National Laboratory under Contract
No. DE-AC05-00OR22725 managed by UT-Battelle, LLC.

\end{document}